\documentclass{sig-alternate-2013}
\usepackage{verbatim}

\pdfoutput=1

\usepackage{booktabs}
\usepackage{array}
\newcolumntype{P}[1]{>{\centering\arraybackslash}p{#1}}

\permission{\copyright 2017 International World Wide Web Conference Committee \\ (IW3C2), published under Creative Commons CC BY 4.0 License.}
\conferenceinfo{WWW 2017,}{April 3--7, 2017, Perth, Australia.}
\copyrightetc{ACM \the\acmcopyr}
\crdata{978-1-4503-4913-0/17/04. \\
http://dx.doi.org/10.1145/3038912.3052666  \\
\includegraphics{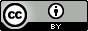}
}

\begin{document}

\title{Discussion quality diffuses in the digital public square}

\numberofauthors{2}

\author{
\alignauthor George Berry\\
       \affaddr{Department of Sociology}\\
       \affaddr{Cornell University}\\
       \email{geb97@cornell.edu}
\alignauthor Sean J. Taylor\\
       \affaddr{Facebook}\\
       \email{sjt@fb.com}
}

\maketitle

\begin{abstract}
Studies of online social influence have demonstrated that friends have important effects on many types of behavior in a wide variety of settings. However, we know much less about how influence works among relative strangers in \emph{digital public squares}, despite important conversations happening in such spaces. We present the results of a study on large public Facebook Pages where we randomly used two different methods---\emph{most recent} and \emph{social feedback}---to order comments on posts. We find that the \emph{social feedback} condition results in higher quality viewed comments and response comments. After measuring the average quality of comments written by users before the study, we find that \emph{social feedback} has a positive effect on response quality for both low and high quality commenters. We draw on a theoretical framework of social norms to explain this empirical result. In order to examine the influence mechanism further, we measure the similarity between comments viewed and written during the study, finding that similarity increases for the highest quality contributors under the \emph{social feedback} condition. This suggests that, in addition to norms, some individuals may respond with increased relevance to high-quality comments.

\end{abstract}


\section{Introduction}

Studies of social influence primarily focus on socially connected individuals \cite{Marsden1993, Christakis2007a, Fowler2008, Bakshy2009, Aral2009b, Centola2010b, Bakshy2012, Coviello2014, Lewis2012}, yet discussions on important issues often occur between relative strangers. In large public online discussions, which we term \emph{digital public squares}\footnote{Examples are large public Facebook Pages and Groups, groups of Twitter users who discuss particular topics, and large sub-Reddits.} \cite{Rhue2014}, understanding social influence processes is crucial for designing systems that encourage meaningful discussions.

In this paper, we present the results from a large-scale study on comments sections from public Facebook Pages. Our experimental change is whether to display comments on posts ranked by \emph{social feedback} (treatment) or \emph{most recent} (control). In contrast to previous work on large online conversations \cite{Zhang2010, Romero2011, Tan2016}, randomization allows us to interpret our findings causally.

Our study makes the following contributions to understanding online discussions:

\begin{itemize}
\item We provide a framework for studying the quality of comments and demonstrate that text-only models can be used to predict or measure quality. In contrast, previous work using text has largely focused on affect or sentiment analysis \cite{Golder2011, Kramer2014, Coviello2014}.

\item We evaluate ranking methods on the dimension of quality shown to users and characterize how showing higher quality can improve user experience. Previous work has focused on implicit quality or predicted feedback \cite{Hsu2009, Dalal2012}.

\item We use pre-treatment information to distinguish between competing hypotheses of selective turnout (i.e. \emph{social feedback} changes who participates) versus within-viewer quality change (i.e. \emph{social feedback} causes changes in how they participate).

\item We provide evidence that ranking affects the social norms operating in an online discussion environment, leading to increased quality and feedback, as well as improved quality of response comments. We additionally find that the \emph{social feedback} condition encourages increased relevance of response comments in certain cases.

\item Methodologically, we demonstrate that within-subjects designs can be used for studying discussions, with the benefit that we do not change ranking for all posts a user sees, or for all users that see a post. This allows ranking methods to be evaluated with minimal user experience change.

\end{itemize}

\noindent
This research was conducted as part of a product test that was designed to evaluate the effectiveness of the comment ranking algorithm employed on large public Facebook pages in accordance with Facebook's research policies\footnote{An explanation of Facebook's research review process can be found in \cite{Jackman2016a}.}. We note that all comments analyzed in this study were posted publicly, and that all information was analyzed anonymously and in aggregate.

\section{Previous work}

We draw on several areas of research in this study, which we summarize here.

Previous research on social influence has established that product adoption, sentiment, political participation, and cultural taste are transmitted through online social networks \cite{Aral2009b, Bond2012, Bakshy2012, Kramer2014, Muchnik2014}. This diffusion generally occurs in the context of a friendship, or where individuals have a prior social relationship. At the same time, studies have established that social influence can operate even in anonymous-interaction cases \cite{Salganik2006, Keizer2008, Tsvetkova2014}, both offline and online. This work provides an expectation that individuals in digital public squares should be influenced by the behavior of others, even if a prior social relationship does not exist.

A growing literature has sought to understand different aspects of large online discussions. From a technical perspective, researchers have investigated methods for representing discussions in latent spaces \cite{Goldberg2014a, Le2014, Vosoughi2016}, ranking comments by inferred community preferences \cite{Hsu2009, Dalal2012}, and modeling the lifecycle of discussion threads \cite{Wang2011a}. We draw heavily on this work in building tools for our study. From a content perspective, researchers have developed notions of quality in online discussions \cite{Otterbacher2009, Diakopoulos2011} and identified patterns of trolling behavior in online forums \cite{Shachaf2010, Cheng2012, Cheng2014}.

Work in social science seeks to understand how social norms affect social behavior, particularly in public spaces \cite{Coleman1994, Schultz2007, Cialdini1990, Cialdini1991}. When different norms are activated in otherwise identical places, different behavioral outcomes are observed\footnote{The classic example is littering. If there are zero or one pieces of litter on the ground, individuals perceive an anti-littering norm and act accordingly. On the other hand, if there are many pieces of litter, individuals adopt a norm indicating that littering is okay, and behave accordingly.}. It has also been found that the violation of one norm (e.g. a norm against littering) can encourage the violation of a separate norm (e.g. a norm against graffiti) \cite{Keizer2008}. This research identifies two important classes of norms: \emph{descriptive} and \emph{injunctive} norms. In a social situation, the former indicates which behavior is most common (e.g. not many people litter here) whereas the latter provides information about which behavior is appropriate (e.g. don't litter here). We primarily focus on descriptive norms in this paper.

Finally, theoretical work in social science and the humanities investigates the social implications of the transition of public discussions to online spaces \cite{Dahlgren2005, Boyd2007, Wojcieszak2009, Zhang2010, Shirky2011}. This research can be traced to \cite{Habermas1991}, which develops the idea of a ``public sphere''---a space where citizens come together as equals to exchange information and persuade one another about issues relating to the common good. Scholars have argued that the Internet has pluralized the public sphere into ``networked publics'' \cite{Boyd2007} organized around foci \cite{Feld1981}, which have properties (e.g. searchability) not found in offline discussions.

\subsection{Research questions}

Recall that our treatment condition is ranking comments by \emph{social feedback}, while our control condition is ranking them by \emph{most recent}. Based on previous work, we developed the following set of questions.

\begin{itemize}
\item Does ordering comments by \emph{social feedback} increase the average comment quality shown to viewers?
\item Does ordering comments by \emph{social feedback} increase the probability that a viewer writes a response comment? Does it increase the probability of other engagement with comments (e.g. likes)?
\item Does ordering comments by \emph{social feedback} increase the average quality of a response comment given a viewer writes one?
\item Assuming there is a change in response comment quality, can this change be attributed to \emph{selective turnout} of certain types of commenters, or a within-viewer \emph{quality change}?
\item Assuming that there is a change in response quality in the treatment condition, are responses under treatment more similar to comments viewed?
\end{itemize}

In this study people who use Facebook can play two roles: comment author and comments viewer.  Authors are the people who wrote the comments that the viewer is currently seeing.  The \emph{viewers} are the subjects of the experiment whose behavior we are studying.  The \emph{authors} are producing the content that we are deciding how to order, providing the different stimuli for the viewers.  In order to clarify this distinction as well as provide formal definitions of some other concepts we measure, we have provided Table~\ref{tab:terms}.

\begin{table}[t]
  \caption{Definition of the two user roles, measurements of the state viewers see, and measurements of the outcomes they produce.}
  \begin{tabular}{P{2.5cm} p{5cm}}
  \toprule
  Term & Definition \\
  \midrule
  Author & An individual who writes a comment\\[3pt]
  Viewer & An individual who sees a comment\\[3pt]
  Viewed quality & The quality of comments viewed\\[3pt]
  Response quality & The quality of comments written\\[3pt]
  Historical quality & Average quality of comments written on public Pages in 14 days before beginning of test\\
  \bottomrule
  \end{tabular}
  \label{tab:terms}
\end{table}

\section{Design}

We conducted a test between June 27, 2016 and August 1, 2016 on large public Facebook Pages. This test was run as part of a product test for evaluating the effectiveness of the Page comment ranking system.

We randomized the method for ranking discussions on these Page posts for viewer-post pairs. If we denote viewers $v$ and posts $p$, then each $(v, p)$ pair had a small chance of being in the test, and if it was included in the test, had an equal chance of being assigned to \emph{social feedback} ranking or \emph{most recent} ranking. This resulted in 45.4 million viewer-post pairs, comprising 25.9 million unique viewers and 6.7 million unique posts. Because our comment quality measurement methodology (described in Section~\ref{sec:quality}) was performed in English, this test was restricted to Facebook users who primarily speak English.

\begin{figure*}[t]
\centering
\includegraphics[scale=0.5]{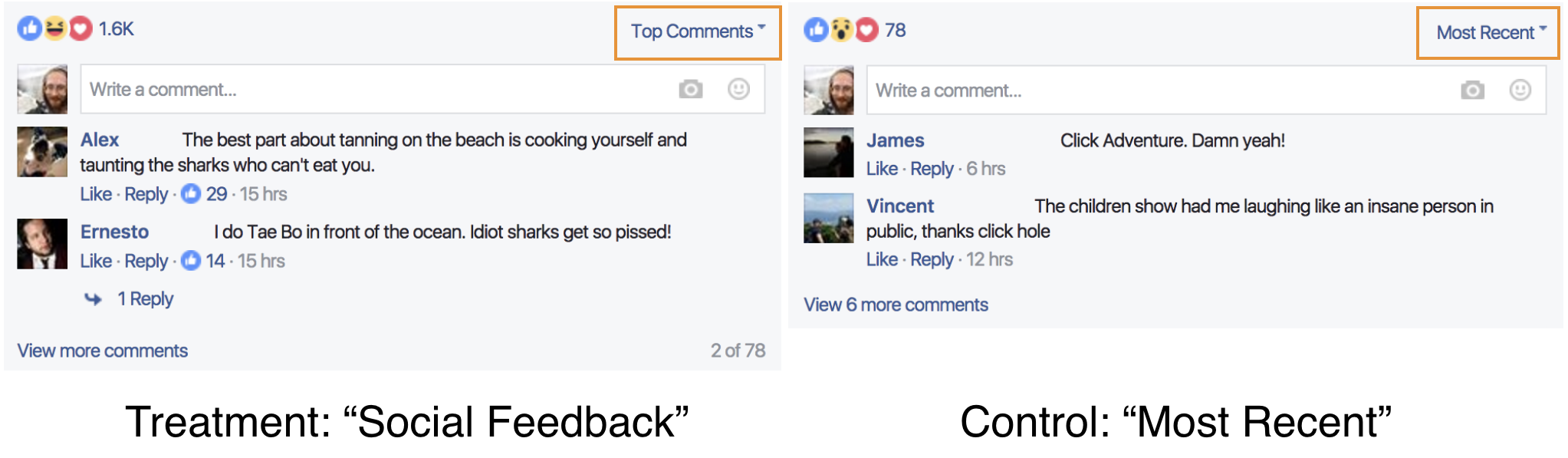}
\caption{The user interfaces for treatment and control conditions that were presented to the viewer during the test. This test was conduced on the Web browser version of Facebook.}
\label{fig:interface}
\end{figure*}

Both \emph{social feedback} and \emph{most recent} are available for viewers to choose on posts made by large Pages via a dropdown element at the top right of the comments section (see Figure \ref{fig:interface}). We changed the default method for presenting post discussions while still allowing viewers the option to choose alternate ranking methods. $(v, p)$ pairs were placed in a single condition for the duration of the test, meaning that repeated viewings of the same post's comments resulted in the same default ranking. In conducting our analyses, we analyzed all data in aggregate, and present only aggregate measures here. In addition, all posts and comments analyzed for this study are from Pages and therefore public information in accordance with Facebook's Data Policy.

\subsection{Randomization}

We chose a \emph{within-subjects} design for several reasons. Importantly, this randomization minimally changes any individual's experience, since only a small number of posts each viewer sees will be in the test. Additionally, viewer-post randomization prevents two potential difficulties in experimental interpretation: \emph{spillover effects} and \emph{inventory effects}.

\emph{Spillover effects} can result from changing several consecutive posts for a given viewer. In this case, assigning post $p_1$ to a category affects the response to $p_2$, for instance by changing the baseline expectations for $p_2$. Randomization at the viewer-level could result in changed expectations across conditions.

We also wanted to avoid \emph{inventory effects}, which could occur if we randomized at the post-level (where all viewers of a particular post see the same ordering method). Here we make a distinction between the effectiveness of a ranking method on a given inventory, and the effectiveness of a ranking method at generating an inventory. Intuitively, a fair test of a ranking method is applying different ranking methods to similar inventories. If different ranking methods generate different inventories, interpreting experimental results from a post-level randomization could conflate two factors: the inventory effect and the ranking method effect.

Randomizing at the viewer-post level avoids these problems and allows us to interpret total effects presented in this paper as the effect of the ranking method itself. In expectation, all posts have the same chance of being displayed according to \emph{social feedback} and \emph{most recent}, and therefore have similar processes generating comment inventories. At the same time, it is very unlikely that there are spillovers at the viewer level because the chance of seeing two consecutive posts in the test is very low.

\section{Measuring discussion quality} \label{sec:quality}

\begin{table}[t]
  \caption{A summary of the rating guidelines for comment quality given to human raters. We consider a comment rated 3 or higher to be ``high quality'', in that it contributes to the discussion.}
  \centering
  \begin{tabular}{P{1cm} p{6.5cm}}
  \toprule
  Rating & Description \\
  \midrule
  1 & Comments that might negatively affect the user experience; including out of context, spam, aggressive language or irrelevant comments.\\[3pt]
  2 & Comments that might be relevant but do not further the conversation.\\[3pt]
  3 & Comments that perpetuate continuance of the conversation; signified by an original composition, question or opinion.\\[3pt]
  4 & Presents multiple ideas, presents new information beneficial to the reader, or brings new perspectives relevant to the post.\\[3pt]
  5 & Makes an in-depth, interesting, engaging statement/question that is worth reading, and adds to the comment conversation in an exceptional or noteworthy way.\\
  \bottomrule
  \end{tabular}
  \label{tab:quality_guidelines}
\end{table}

Our measure of quality is focused on whether a comment adds to a conversation. We operationalize this as a measure of quality judged by human raters in the context of a post, according to guidelines seen in Table \ref{tab:quality_guidelines}\footnote{Our raters have performed this rating task professionally for an extended period of time and were not new to this task.}. We chose a ``know it when you see it'' description of quality because we wanted our measure to reflect the raters' intuitive judgments\footnote{The alternative would have been to specify a set of specific elements (e.g. comment contains a question). Because we do not know how specific elements map to quality, we chose to rely on intuitive ratings rather than an artificial set of elements that together make up ``quality''.}.

In addition, latent quality is often the objective of recommender methods. In essence, we are asking the raters how specific comments align with their latent notion of quality. This method also has the upside that a sufficiently rich representation of comment text, combined with enough training data, can allow us to learn the important elements of quality from the data.

To build our rated set of comments, we sampled from the 5000 largest English-language Facebook Pages, which include posts and discussions on a wide variety of topics. We asked 23 raters to evaluate a sample of 100,000 comments\footnote{Comments were sampled in proportion to how many were viewed by viewers, so the are not representative of comments written by Facebook users, but of the comments Facebook users see on pages.} that were predicted to be written in English by a language classifier.  Inter-rater reliability was fair (Krippendorff's $\alpha$ = 0.714) and each comment was labeled by 2 raters so we could use only comments for which they agreed. We used the consensus quality labels to build a classifier that predicted the quality of a comment from its text only.

To simplify the prediction problem, we binarized the outcome measure.\footnote{Ratings of 1, 4, and 5 were relatively rare compared to 2 and 3, so we believe very little information has been discarded in this transformation.} If the rating was 3 or higher---corresponding to a comment contributing to the discussion---we assigned the comment the label of ``high quality'', otherwise we assigned it the label of ``low quality''. This lead to a prediction problem where 34.6\% of comments were ``high quality'' and the remaining comments were ``low quality''. We built two separate models (see Appendix) and achieved an area under the ROC curve (AUC) measure of 0.85 for the first model and 0.90 for the second. All results presented here are using the second model to predict quality.

\section{Results}

In this section, we first describe the outcome variables we measure for the test.  We then start with average treatment effects, then study two types of effect heterogeneity---by number of available comments and pre-treatment commenter quality.  Finally we attempt to unpack the mechanism for the quality effect in the last subsection.

\subsection{Outcome Variables}

We measure the following variables as outcomes in our test:

\begin{itemize}
\item \emph{Response comments}: The number of response comments written by viewers of the post.  Although we call them ``response'' comments, they may not necessarily be replies to existing comments.
\item \emph{Response likes}: The number of likes by the viewer on any comment they see on a post in the test.
\item \emph{Viewed quality}: The average model-predicted quality of the most highly ranked two or more comments viewed.
\item \emph{Response quality}: The quality average model-predicted quality of the response comments written by the viewer.  This measure is only available in the case that the viewer writes at least one response comment.
\item \emph{Response similarity}: The cosine similarity between the text of the most highly ranked comments the viewer saw and response comments they wrote in response.  As with \emph{response quality}, this measure is only available in the case that the viewer writes at least one response comment.
\end{itemize}

\subsection{Average effects} \label{sec:basic}

\begin{figure*}[h]
\centering
\includegraphics[width=0.8\textwidth]{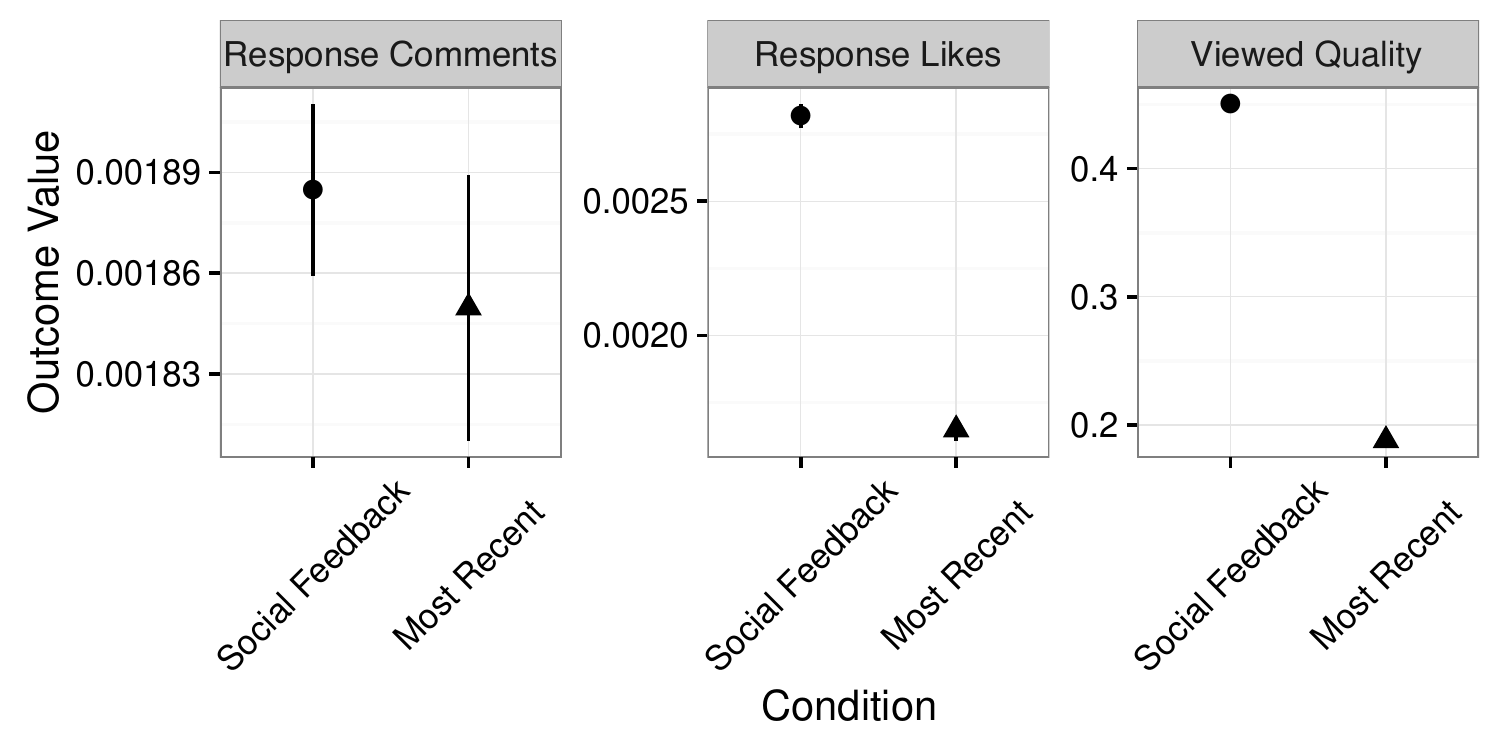}
\caption{Average treatment effects of \emph{social feedback} (circles) compared to \emph{most recent} (triangles) on the number of comments written, the number of likes on post comments, and the predicted quality of viewed comments.}
\label{fig:basic_ate}
\end{figure*}

Figure \ref{fig:basic_ate} presents average treatment effects for three measures: the number of response comments, the number of likes, and the predicted quality of viewed comments. Relative to the \emph{most recent} condition, \emph{social feedback} increases predicted quality of viewed comments and the number of response likes. An increased number of likes indicates that individuals are finding a more worthwhile reading experience in comment sections ordered by \emph{social feedback}. Interestingly, the number of response comments does not increase significantly in the \emph{social feedback} condition. This suggests that choices to comment are more stable with respect to displayed quality than liking behavior.

Importantly, the predicted quality of shown comments is increased from 0.2 to 0.45 in the \emph{social feedback} condition. This large change suggests that taking viewer feedback into account leads to displaying much higher quality comments, which in turn suggests that platforms can use ranking to create higher quality experiences for viewers. The magnitude of this change is large, but we note that a particular feature of the Facebook platform makes it less surprising: people often tag a friend in public comments sections as a method of alerting that friend to interesting content\footnote{For instance, a comment may simply say ``John Smith'', which will then send a notification to John pointing to the post.}.

Such comments are unlikely to be liked by many people, and unlikely to be classified as high quality comments. In this way, \emph{social feedback} performs a type of collaborative filtering to de-emphasize comments that are only directed towards a single individual and not the discussion as a whole.

In addition to these effects, we also find that \emph{social feedback} increases the relative risk of a response comment being high quality by 50\% ($t = 61$), indicating that comments written in response to the treatment condition are much more likely to be classified as high quality.

Finally, we find that the similarity of response comments to viewed comments decreases by 5\% ($t = 6.6$) in the \emph{social feedback} condition relative to the \emph{most recent} condition. However, as discussed below, this average treatment effect exhibits strong heterogeneity.

\subsection{Effects by inventory}

\begin{figure}[h]
\centering
\includegraphics[width=0.4\textwidth]{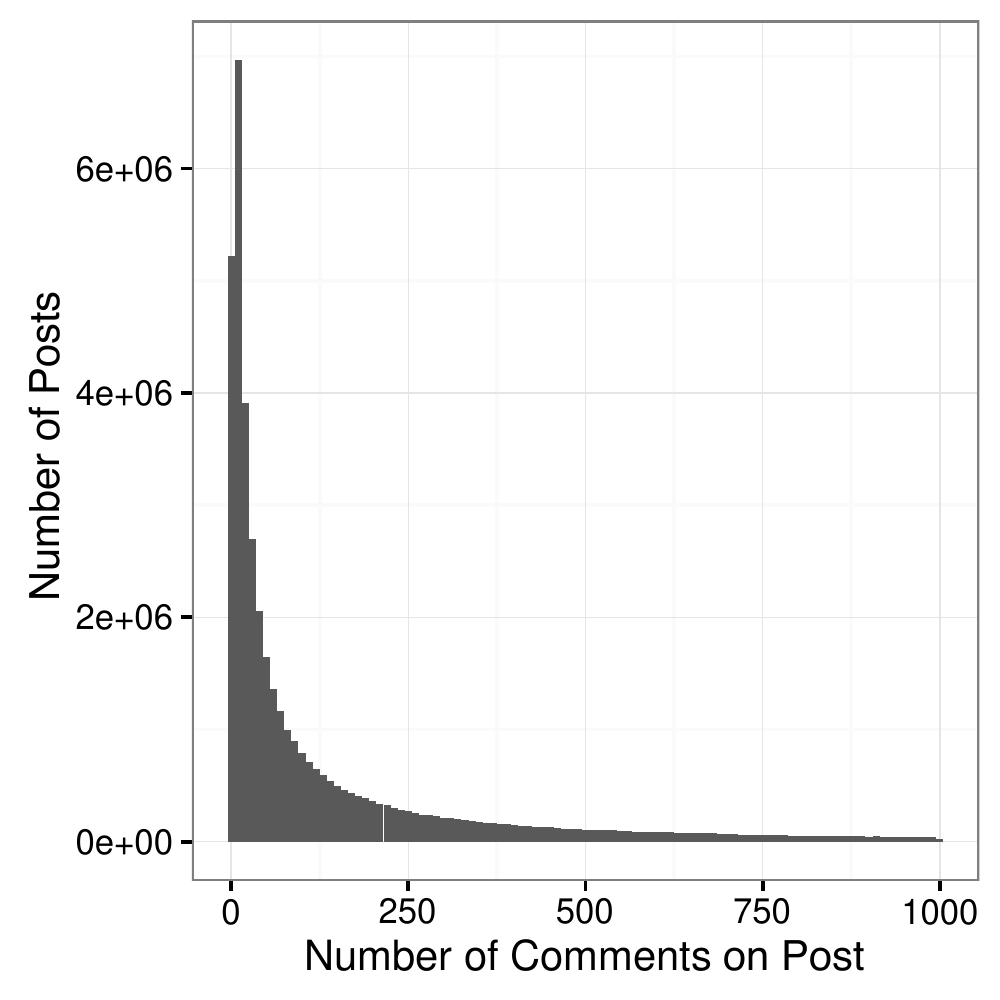}
\caption{The distribution of the number of available comments on posts in the study, truncated at 1000 in order to make it more readable.  There is a long tail of posts with much larger numbers of comments.}
\label{fig:inventory_distribution}
\end{figure}

\begin{figure*}[h]
\centering
\includegraphics[width=0.8\textwidth]{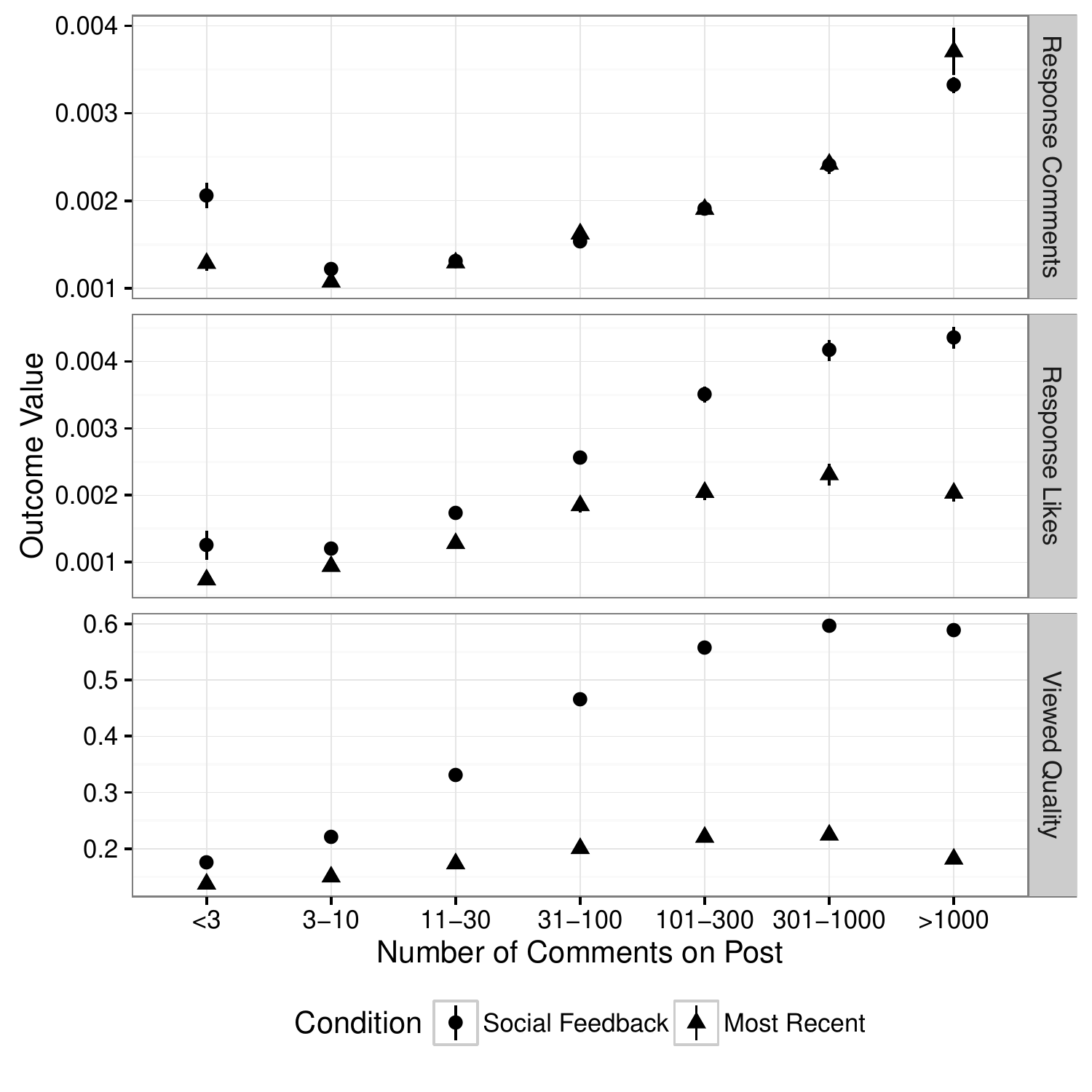}
\caption{Treatment effects presented by post comment inventory. From top to bottom: response comments written by the viewer, response likes on post comments, and viewed comment quality.}
\label{fig:cate_by_inventory}
\end{figure*}

The number of comments available to rank is an important factor elided by the average treatment effect plots in Figure \ref{fig:basic_ate}. Intuitively, if there is a larger comment inventory, ranking methods should have larger effects. Despite being the default on the Web, ranking by \emph{most recent} is somewhat unique in that it does not choose from a pool, but simply displays the last comments to be written. This implies that as inventories get larger, we should observe larger differences between \emph{social feedback} and \emph{most recent}.  Figure~\ref{fig:inventory_distribution} shows the distribution of comments available to rank, or inventory, when the viewers in the test first saw the posts.

Stratifying the plots in Figure \ref{fig:basic_ate} by comment inventory produces Figure \ref{fig:cate_by_inventory}. In the case of likes and shown quality, we find that an increasing inventory is associated with larger differences between conditions.

Interestingly, up until discussions reach 1000 comments, likes and shown quality increase in the control case, and \emph{social feedback} widens the gap between the treatment and control cases. This suggests that there are basic dynamics of conversations present under \emph{most recent}: as discussions grow larger, they become higher quality. \emph{Social feedback} ranking accelerates this process, quickly leading to higher quality, more liked discussions.

For very large discussions, defined as those reaching over 1000 comments, we see slightly different dynamics. Compared to discussions with 301 to 1000 comments, likes and shown quality decrease in the \emph{most recent} condition, while the the number of response comments increases substantially. This suggests that when discussions grow very large, there is a pile-on effect of many individuals writing low-quality comments---perhaps tagging friends to alert them to a post of particularly broad appeal.

For very large discussions displayed by \emph{social feedback}, the increase in quality levels off, relative to the 301-1000 group. This could indicate that rich-get-richer effects \cite{Salganik2006} have resulted in certain comments getting a huge number of likes, preventing new, high-quality comments from rising to the top.

\subsection{Selective turnout versus quality improvement}

\begin{figure*}[h]
\centering
\includegraphics[width=0.8\textwidth]{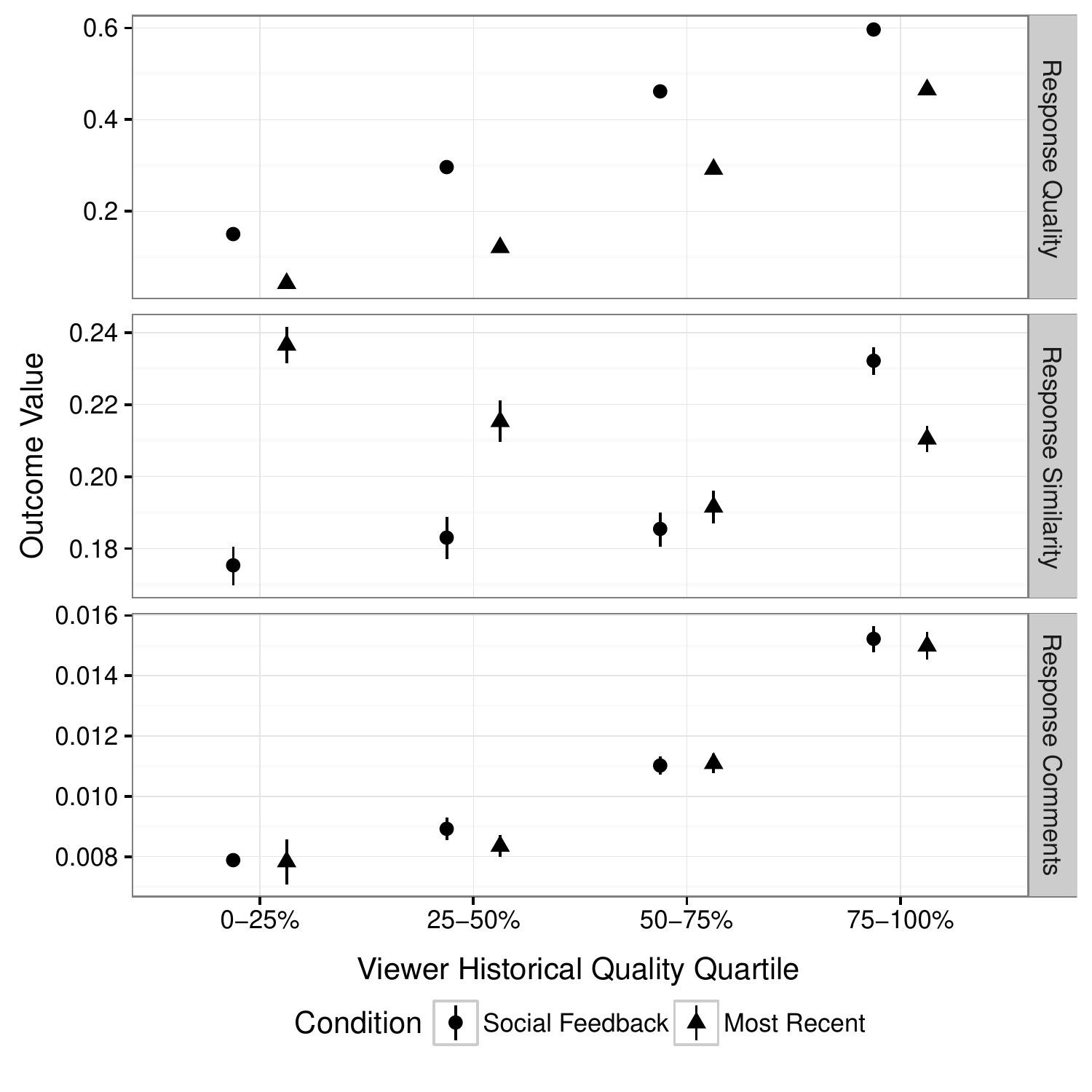}\caption{Treatment effects presented by quartile of historical quality of the viewer. From top to bottom: the quality of response comments written by the viewer, the similarity between viewed and response comments, and the number of response comments written. Only individuals with at least 10 comments in the 14 days prior to the test are included.}
\label{fig:cate_by_quality}
\end{figure*}

We can explain the total effect of increased response quality through at least two distinct social mechanisms: the treatment encourages high quality commenters to turn out at a higher rate (\emph{selective turnout}), or it encourages a fixed group of commenters to write higher quality comments (\emph{quality change}) \cite{Taylor2013a}. In a given setting, both of these mechanisms can operate, making interpretation of an average effect difficult.

Selective turnout and quality change have different implications for understanding how a platform operates. For instance, if the increase in response quality is driven entirely by selective turnout, a change that appears to have a positive effect may shrink the set of viewers that ultimately participate in discussions. This undesirable result could be masked by only considering average treatment effects. If the changes are entirely due to increased within-viewer response quality, then the same concerns about affecting turnout in a negative way do not apply. We can also imagine many intermediate states: for instance, perhaps only a turnout effect exists, but high quality commenters write more while nobody writes less.

Empirically, using only the data from the test, it is impossible to directly disentangle these two effects without making additional assumptions\footnote{See the potential outcomes table in the appendix of \cite{Taylor2013a}.}. This results from the treatment affecting two decisions: whether to comment, and what quality to write. We never observe all potential outcomes. For example, if the treatment causes me not to comment, what would my quality have been if I did comment? We cannot answer this question directly.

\begin{figure}[h]
\centering
\includegraphics[height=0.5\textwidth]{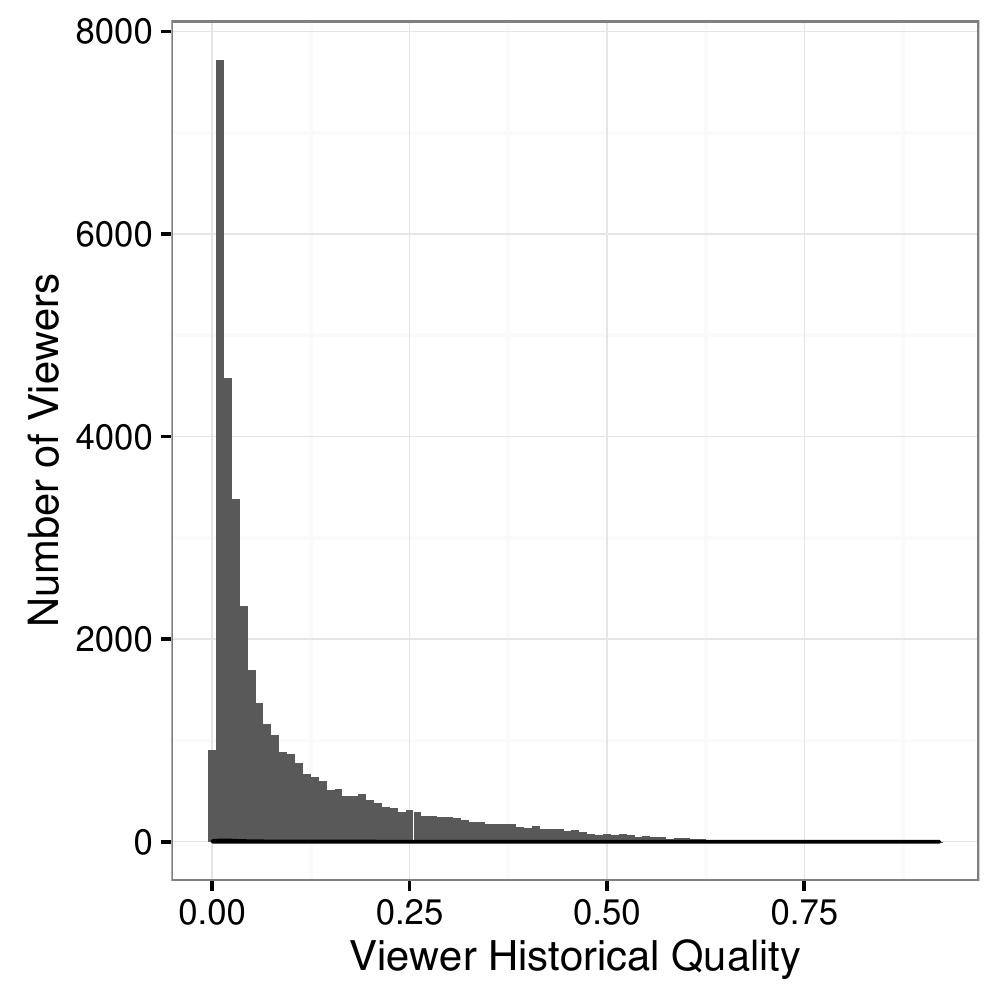}
\caption{Distribution of Average Comment Quality}
\label{fig:quality_distribution}
\end{figure}

However, if we have access to pre-treatment information on commenter quality (\emph{historical quality}), we can examine conditional average treatment effects. We define historical quality as the average quality written on public Page posts in the 14 days before the test. We limit this analysis to individuals who had at least 10 such comments in this period. We present the distribution of this measure in Figure~\ref{fig:quality_distribution}: it is relatively skewed and consistently high-quality commenters are rare.

Figure \ref{fig:cate_by_quality} presents response quality, response similarity, and number of response comments, all conditioned on historical quality. Across all levels of historical quality, we find that \emph{social feedback} ranking increases response quality. This indicates that there is a relatively uniform quality change effect. On the other hand, we see that \emph{social feedback} has little effect on the number of comments written at any level of historical quality. This indicates the lack of a substantial turnout effect.

These two findings together suggest that the average treatment effect presented in Section \ref{sec:basic} results primarily from \emph{quality change} rather than \emph{selective turnout}. We qualify this by noting we have chosen to condition only on historical quality, and that conditional average treatment effects may look different when conditioning on other variables. However, if we assume that historical quality is a stable feature of individuals, the interpretation is clear: \emph{social feedback} ranking increases response quality across the spectrum while leaving comment probability virtually unchanged.

Since the control condition, \emph{most recent}, is the default method for presenting conversations on the Web, this result suggests that alternate ranking methodologies can substantially increase discussion quality among a fixed set of commenters. Put another way, if we view participants in social media discussions as having a ``comment budget,'' these results suggest that ranking can encourage higher quality discussions even if budgets for everyone remain fixed.

We find a broad quality change effect, which implies that the \emph{social feedback} condition makes salient a descriptive social norm for high-quality commenting. We argue that a norm is the most reasonable explanation for the observed diffusion of quality, because it operates across a broad range of Pages and commenters. In addition, individuals do not have an existing social connection with the authors of the comments they read. In such situations, individuals look to strangers to gather information about common and successful behavior \cite{Keizer2008, Peysakhovich2015}.

\subsection{Response comment similarity}

Finding that the ranking method changes salient descriptive norms does not indicate that no other factors are at work. One such element is the \emph{similarity} of response comments to viewed comments, which we define using cosine similarity and a TF-IDF representation of comments. If \emph{social feedback} only makes salient a descriptive norm of high-quality commenting, then we would not expect changes in the relevance of response comments to the ongoing discussion.

Results for this analysis are presented in Figure \ref{fig:cate_by_quality}, middle panel. An interesting pattern emerges, which clearly indicates that the ranking method affects the similarity of response comments. However, the effect is not uniform across historical quality levels. For the lowest quartile, response comments are less similar in the \emph{social feedback} condition. For the highest quartile, response comments are more similar. A possible explanation for this finding is that higher quality commenters engage directly with viewed comments, adopting similar language. On the other hand, lower quality commenters may introduce more novelty into the discussion by using terms that are different from those viewed.

This analysis also shows that ranking methods can affect factors of comments apart from quality. Not all factors move uniformly across the historical quality distribution. While all individuals are inclined to write higher quality comments in the \emph{social feedback} condition, these comments have varying degrees of similarity to viewed comments.

\section{Conclusion}

We have shown that the quality of discussions diffuses in digital public squares. We find this result surprising because individuals generally do not personally know one another. This implies that improving ranking methods has both first-order and second-order effects. Ranking improves the reading experience by displaying higher quality content for readers. Then, for the small fraction of individuals that choose to participate in the discussion, displaying better comments encourages higher-quality participation. We attribute this higher quality participation largely to descriptive norm adoption.

While we see our results as encouraging, there are several limitations that should be addressed in future research. First, we have used an omnibus measure of quality that can be decomposed into several dimensions. It is possible that not all dimensions of quality diffuse equally. Second, we have simplified the notion of discussion quality by assuming that the quality of the discussion is a sum of its parts. While this is a first-order approximation to discussion quality, important higher-order elements such as opinion diversity are not included in our measure. Third, the extent to which our findings generalize to other settings is an open question. Particular features of the Facebook platform (e.g. personal photos next to comments) may facilitate the diffusion of discussion quality at an increased rate compared to truly anonymous discussion forums. Fourth, we have presented here evidence for a more-or-less dyadic form of diffusion (from the author of a viewed comment to that viewer responding with higher quality). Future research can address the specific effect this diffusion has on the overall outcome of discussion threads. Because diffusion happens at the microlevel, we should expect an overall effect on discussion threads, however a variety of aggregation processes can produce different macrolevel outcomes.

Finally, we suggest that our findings can be applied in a variety of settings. Most naturally, discussion forums---which traditionally present discussion by \emph{most recent}---can consider employing ranking to boost the vibrancy of discussions. Particularly when such discussions are about important issues, we think that presenting the best elements of a discussion is important. In another domain, our results suggest that during the academic peer review process, showing reviewers examples of high quality peer reviews may increase the quality of feedback provided.

\section{Acknowledgments}

We thank Michael Macy, Matthew Brashears, Alexander Peysakhovich, Ana Franco, Dean Eckles, Nan Li, Benjamin Cornwell, Christy Sauper, Monica Lee, Eytan Bakshy, Jenny Wang, and participants in the 2015 Conference on Digital Experimentation for helpful discussion and suggestions. We thank Erich Owens and Ashoat Tevoysan for lending their time and talent to help realize this work. We thank an anonymous reviewer for helpful comments and suggestions.

\appendix

\section{Quality classification}

We built two separate quality classifiers. Both had similar results. Work presented in the body of the paper uses the second model, which performed slightly better on our main evaluation criterion: the area under the receiver operating characteristic curve (AUC for short). We choose this metric because it summarizes two important measures, the true positive rate (TPR) and false positive rate (FPR), along with providing a measure of the tradeoff between them. The AUC has a max of 1 (perfect predictive ability) and a min of 0 (no predictive ability).

The classification task was predicting whether a comment was high quality (1) or not (0), using only its text to generate features.

\subsection{Na{\"i}ve Bayes + GBDT}

We used a two-step modeling procedure. The first step took sparse text features (n-grams) and summarized them into one dense feature using a Na{\"i}ve Bayes classifier. The second step took the output of the first step plus a variety of hand-chosen features and generated our final prediction for the comment.

For the Na{\"i}ve Bayes model, we took a comment's text and did the following to generate word unigrams and bigrams:

\begin{enumerate}
\item Inserted spaces around punctuation
\item Split on spaces
\item Removed common stopwords
\item Generated unigrams and bigrams of remaining tokens
\end{enumerate}

We then returned to the raw text and took unigrams and bigrams of all characters. For instance, in the word ``the'', the character-level unigrams are $(t, h, e)$ while the character-level bigrams are $(th, he)$.

We applied a chi-squared feature selection procedure to this large space of word and character unigrams and bigrams to eliminate features with low predictive power. We then used the surviving features to predict the outcome (high quality comment or not). Since a Na{\"i}ve Bayes classifier outputs a probability of a comment being high quality, we generated a probability of being high quality for each comment based on the first-step model for use in the second-step model.

We then generated about 20 ``hand-made'' features from reading hundreds of comments with quality label annotations. These features included:

\begin{enumerate}
\item Whether a link was present
\item The number of links
\item Whether a mention was present
\item The number of mentions
\item Whether a vulgar word was present
\item The number of vulgar words
\item Whether a smiley or emoji was present
\item The number of smiley or emojis
\item Whether an all-caps word was present
\item The number of all-caps words
\item The number of all-caps letters
\item The fraction of all-caps letters
\item The presence of excessive punctuation (e.g. "........")
\end{enumerate}

We included transforms of these variables as well (squared, etc.). We put these hand-made features together with the probabilistic output of the first-step model. Using these combined features, we predicted the quality of a comment using a gradient-boosted decision tree (GBDT) modeling procedure. We iterated through 50 and 100 trees with depths of 3, 4, and 5. The best model produced an AUC of 0.85.

\subsection{Multilayer perceptron}

The two-step model above is knowledge-intensive, requiring us to learn about our domain to generate features. In addition, it requires cross validation to select the right parameters for several elements, such as feature selection, tree depth, etc.

A simpler approach exists: generating features by embedding text in a latent space using Word2Vec, then using a single model on comment vectors. This approach requires selecting a dimensionality to represent comments in, plus choosing a model to relate comment vectors to quality. However, it is substantially simpler than the procedure presented above because it does not require us to develop a theory of ``what makes a comment high/low quality''. Such a theory could be quite useful in many domains, but for the task at hand we wanted maximum confidence that we were predicting comment quality well.

We chose a dimensionality of 300 for the Word2Vec procedure. This means that each word in the comment corpus was embedded in a 300-dimension space, yielding a 300-vector. For each comment, we took the vectors corresponding to each word in the comment and averaged them to get a comment vector. For instance, if a comment reads ``The dog is very good'', we would take the vectors for ``The'', ``dog'', etc. Then we average those vectors and obtain a 300-dimension comment vector.

To relate these 300-vectors to our output variable (high quality or not), we chose a multilayer perceptron model with 2 layers. This is one of many models we could have chosen, but had the benefit of being straightforward technically and achieving good results. The AUC of this modeling procedure was 0.90.

\bibliographystyle{abbrv}
\bibliography{fbbib1}

\end{document}